\documentclass[amsmath,amssymb,aps,superscriptaddress,nofootinbib,floatfix]{revtex4-2}

\usepackage[T1]{fontenc}
\usepackage{amsfonts,amssymb,stmaryrd,latexsym,amsmath,braket}
\usepackage{float}
\usepackage{graphicx}
\usepackage[export]{adjustbox}
\usepackage{array}
\usepackage{times}
\usepackage{bbm}
\usepackage{bm}
\usepackage{mathrsfs}
\usepackage{slashed}
\usepackage{braket}
\usepackage{hyperref}
\usepackage{verbatim}
\usepackage[utf8]{inputenc}
\usepackage{tabularx}%, tabularray}
\usepackage{siunitx}
\usepackage{xparse}
\usepackage{makecell}
\usepackage{graphicx}
\usepackage{slashed}
\usepackage{braket}
\usepackage{verbatim}
\usepackage{multirow}
\usepackage{booktabs}
\usepackage[utf8]{inputenc}
\usepackage{physics}
\usepackage{amssymb}
\usepackage{graphicx}
\usepackage{mathtools}
\usepackage{xcolor}
\usepackage{xspace}
\usepackage{hyperref}

\newcommand{\FPM}[1]%
{\begingroup{\color{red} (fpm: #1)}\endgroup}

\newcommand{\affAgder}{Department of Information and Communication Technology, University of Agder, N-4604 Kristiansand, Norway}
\newcommand{\affUSN}{Department of Science and Industry systems, University of South-Eastern Norway, N-3616 Kongsberg, Norway}
\newcommand{\affEeroQ}{EeroQ Corporation, Chicago, IL 60651, USA}
\newcommand{\affOslo}{Department of Physics, University of Oslo, N-0316 Oslo, Norway}
\newcommand{\affMSUphys}{Department of Physics and Astronomy, Michigan State University, East Lansing, MI 48824, USA}
\newcommand{\affMSUchem}{Department of Chemistry, Michigan State University, East Lansing, MI 48824, USA}

\begin{document}

\title{Electrons on  Helium and Entangled Quantum Sensors for Particle Physics}

\author{Maria Elena Perruzza}
\affiliation{\affAgder}
\affiliation{\affUSN}

\author{Niyaz R.~Beysengulov}
\affiliation{\affEeroQ}

\author{Stian D.~Bilek}
\affiliation{\affOslo}

\author{Antoine Y.~M.~C.~Camper}
\affiliation{\affOslo}

\author{Jonas B.~Flaten}
\affiliation{\affOslo}

\author{Morten Hjorth-Jensen}
\email{mhjensen@uio.no}
\affiliation{\affOslo}

\author{Gunnar F.~Lange}
\affiliation{\affOslo}

\author{Oskar Leinonen}
\affiliation{\affOslo}

\author{Jan Malamant}
\affiliation{\affOslo}

\author{Francesco P.~Massel}
\affiliation{\affUSN}
\affiliation{\affOslo}

\author{Johannes Pollanen}
\affiliation{\affMSUphys}
\affiliation{\affEeroQ}

\author{Heidi Sandaker}
\affiliation{\affOslo}

\author{Zachary J.~Stewart}
\affiliation{\affMSUchem}

\author{Viktor Svensson}
\affiliation{\affOslo}

\author{Jared D.~Weidman}
\affiliation{\affMSUchem}

\begin{abstract}
Quantum sensors that harness quantum coherence and entanglement are emerging as powerful tools in many fields, including particle physics, promising unprecedented sensitivity beyond classical detection methods. At the same time, electrons trapped on the surface of liquid helium have emerged as a promising quantum computing, and possibly sensing, platform owing to a nearly impurity-free environment and large predicted coherence times. In this context, single-electron confinement and control using microfabricated traps on helium has been experimentally demonstrated, highlighting the feasibility of scalable qubit architectures on this platform. Leveraging these features, and in line with CERN’s Detector R\&D Initiative 5 (DRD5) \cite{drd52024} on Quantum Sensors, we propose here a sensor concept that uses an entangled pair of electron qubits on superfluid helium for particle physics experiments. We outline the motivation for such spatially and spin-entangled sensors, develop the theoretical formalism for two  electrons and their spins and spatial degrees of freedom in a helium-based double-well trap (analogous to a double quantum dot in semiconductor systems), and discuss the potential advantages for detecting rare high-energy events with quantum-enhanced sensitivity. By exploiting quantum entanglement between the two trapped electrons, this sensor concept can surpass classical sensitivity limits, potentially enabling the detection of signals beyond the reach of classical detectors.
\end{abstract}

\maketitle
\section{Introduction}

Rapid advances in quantum technology are opening new avenues for particle detection. High-energy physics experiments, from collider detectors to cosmic observatories, demand extreme sensitivity to rare or weak signals, yet traditional sensor technologies (scintillators, solid-state trackers, calorimeters) are approaching their practical limits in resolution and noise. Quantum sensors offer a path beyond these limits by exploiting intrinsically quantum phenomena such as superposition and entanglement~\cite{degen2017quantum}. Recognizing this potential, CERN's DRD5 (RD quantum) collaboration~\cite{drd52024} is actively exploring novel quantum sensor concepts for both high- and low-energy physics, spanning platforms that include atomic and nuclear clocks, ion traps, superconducting devices, and spin-based systems.

Among spin-based platforms, quantum dots (QDs), nanoscale semiconductor structures that confine electrons in all three dimensions to produce atom-like energy levels, have long been studied as potential detector materials~\cite{esat2024quantum}. Originally developed for optoelectronics and imaging~\cite{Li2023,garcia2021semiconductor}, they are known for fast scintillation responses~\cite{dropiewski2020ultrafast,kang2011cdte}, narrow emission lines~\cite{loskutova2025quantum}, and spectra that are tunable through their size and composition~\cite{chan1998quantum}. These traits have motivated proposals to embed QDs in particle detectors~\cite{choudhury2025}, for instance using so-called ``chromatic'' scintillation signals in calorimeters to improve energy resolution and particle identification. As a solid-state platform, QD spin qubits combine gate-tunable confinement, mature semiconductor fabrication, high-fidelity control, and relatively long coherence times (from tens of milliseconds to seconds in isotopically purified silicon).

In a complementary approach, electrons floating on the surface of liquid helium have emerged as a promising alternative platform~\cite{dykman2025}. A single electron is trapped just above a cryogenically cooled helium film, forming a low-dimensional system with weak coupling to its environment. Its vertical motion is quantized by the helium image potential, producing a Rydberg-like ladder of discrete levels analogous to those of a QD~\cite{monarkha2013two}. Crucially, the absence of a solid lattice implies extremely low noise and very long coherence times~\cite{platzmanQuantumComputingElectrons1999,lyonSpinbasedQuantumComputing2006}: each electron is essentially isolated in vacuum and experiences negligible spin--orbit or hyperfine decoherence~\cite{schusterProposalManipulatingDetecting2010}, leading to predicted electron spin coherence times on the order of $100$~seconds or more~\cite{lyonSpinbasedQuantumComputing2006}. Rather than producing scintillation light, an electron on helium can be coupled to superconducting circuits or radio-frequency readout devices, converting minute changes in its quantum state into measurable signals~\cite{castoria2025,koolstra2026strong}. Single-electron confinement and control using microfabricated, CMOS-compatible traps has recently been demonstrated~\cite{bradburyEfficientClockedElectron2011,castoria2025selective}, so that quantum dots and helium-surface electrons offer complementary strengths: semiconductor integration and tunability on the one hand, and an almost dissipation-free vacuum environment on the other.

Beyond any single-device readout, our central interest is quantum-enhanced sensing using entangled electrons on helium~\cite{beysengulov2024,leinonen2025}. Each surface-bound electron acts as a qubit sensor, and the key idea is to harness entanglement among multiple electron spins (or motional states) to boost sensitivity. When sensors are entangled, their measurement outcomes become correlated in ways that can outperform unentangled sensor arrays: an entangled network of $N$ qubit sensors can in principle reach a phase sensitivity scaling as $1/N$, approaching the Heisenberg limit, whereas $N$ independent sensors are bounded by the standard quantum limit $1/\sqrt{N}$~\cite{giovannetti2004quantum,giovannetti2006quantum,degen2017quantum}.

A particularly powerful strategy is {\it differential} sensing. Consider two spins prepared in an entangled singlet state (total spin zero). This state is insensitive to uniform, common-mode fields acting on both spins, but is extremely sensitive to any difference between their local environments. If a particle-physics event (for example a charged particle, or a transient field from a rare interaction) produces a small magnetic or electric perturbation at one spin site but not the other, the entangled state acquires a phase or partially decoheres, which can be read out by joint spin measurements. The entangled pair thus acts as a noise-canceling differential sensor: common-mode backgrounds (uniform fields, vibrations) are rejected, while a localized signal, even a minute one, produces a clear change in the spin correlations. Both QD spins and helium-bound electron spins can be operated in this way.

In practical terms, an entangled array of electrons on helium could register tiny energy transfers far below the kiloelectronvolt scale that typically triggers scintillators, semiconductor calorimeters, or tracking detectors. Potential targets include sub-eV energy deposition from light dark matter~\cite{essig2016direct,schutz2016detectability}, single-electron or Rydberg-level excitations, and the extremely weak, slowly oscillating fields associated with ultralight dark-matter candidates such as axions or axion-like particles~\cite{graham2013new,abel2017search,budker2014proposal}, which would couple weakly to electron spins as effective time-dependent Zeeman shifts. Networks of entangled quantum sensors (optical, mechanical, or spin-based) have indeed been proposed to enhance sensitivity to dark-matter and axion signals beyond classical detector arrays~\cite{brady2022entangled,derevianko2018detecting}. Concretely, an array of entangled spin pairs (using either QDs or electrons on helium) embedded in or around a particle detector could serve as a trigger or complementary sensor for new phenomena. For example:

\begin{itemize}
\item \textbf{Ultrafast transients.}  A passing exotic particle, for example a milli-charged particle (hypothetical particles arising in extensions of the standard model that carry an electric charge much smaller than the elementary charge $e$) or hidden-sector boson (such as dark photons, axions or axion-like particles), could produce a rapid, localized magnetic or electric pulse on a microsecond or faster timescale.  Entangled spin pairs, with their long coherence time, could capture this transient: a short pulse at one spin would break the entanglement, signaling the event.  Standard electronics often struggle with such brief spikes, but a quantum sensor's phase coherence acts as a memory of the perturbation.

\item \textbf{Rare event detection.} In searches for very rare processes (light dark matter scattering, electric dipole moments, etc.), an entangled spin ensemble can function as a high-precision magnetometer or calorimeter at the quantum limit.  A small energy deposition or magnetic field change from a rare scattering would alter the spins' coherence.  In practice, one could load many entangled spins in a low-noise environment; if one or a few spins experience decoherence, the correlation signal across the array flags the event.
\end{itemize}

Entanglement-enhanced metrology has already been demonstrated in photonic and atomic systems, where precision beyond the standard quantum limit has been achieved~\cite{aasi2013enhanced,wasilewski2010quantum,pezze2018quantum}. In solid-state platforms, entangled networks are also beginning to appear: ensembles of nitrogen-vacancy (NV) center spins in diamond have been entangled to improve signal-to-noise ratios and to measure tiny magnetic fields well below the classical noise floor~\cite{neumann2008multipartite,wu2025spin,zhou2025entanglement,Sekiguchi2024}. Theoretical work on entangling electrons on liquid helium is likewise an active area~\cite{mostameQuantumSimulatorIsing2008,zhangJaynesCummingsModelsTrapped2009,beysengulov2024,leinonen2025}, and experimental control over single and multiple electrons is improving rapidly~\cite{castoria2025,koolstra2026strong}. Both platforms share practical constraints that any detector application must address: they require cryogenic operation (milli-kelvin temperatures for QD spins, and sub-kelvin cooling for electrons on helium to maintain a stable superfluid film and long spin lifetimes~\cite{monarkha2013two,lyonSpinbasedQuantumComputing2006}), and their radiation hardness under sustained high-energy flux remains to be characterized, although superfluid helium is an inert, defect-free substrate free of stray charges and nuclear spins. A network of spin-entangled electrons on helium would combine the extraordinary coherence of the helium surface environment with the precision gains of entanglement, potentially revealing subtle signatures of new physics that would otherwise be masked by noise in conventional detectors.

In this article we present a theoretical proposal for an entangled spin sensor based on electrons on liquid helium, with some parallels to quantum-dot implementations. We first describe the relevant properties of electrons on helium as a detector medium and develop the formalism for an entangled electron-spin system, showing how a pair, or network, of such spins could function as a highly sensitive detector. We then quantify the attainable precision through the quantum Fisher information and the quantum Cram\'er--Rao bound, and finally discuss how the scheme might be implemented in practice, its potential advantages for particle detection, and the main challenges, such as decoherence, readout, and integration into an experiment.

\section{Results with an entanglement-enhanced qubit sensor framework}
\label{sec:formalism}

In this section we describe our main results for the chosen two-electron device and develop the formalism that turns it into an entanglement-enhanced sensor. We first specify the confining potential and diagonalize the single-particle problem to obtain well-localized orbitals (Secs.~\ref{sec:device}--\ref{sec:spectrum}). We then construct the two-electron states by configuration interaction, identify the maximally entangled $(1,1)$ singlet that serves as the sensing resource, and quantify its entanglement (Sec.~\ref{sec:cientangle}). We next show how an external perturbation is imprinted on this entangled state through coherent singlet--triplet mixing (Sec.~\ref{sec:signal_encoding}), and how the attainable precision is bounded by the quantum Fisher information and the quantum Cram\'er--Rao bound (Sec.~\ref{sec:qfi-est}). Throughout, emphasis is placed on measurement strategies that are compatible with interaction-generated entanglement and that preserve the noise resilience intrinsic to the correlated spin states. The treatment parallels that of spin-entangled semiconductor double quantum dots~\cite{degen2017quantum,petta2005coherent,burkard2023semiconductor}, but is here realized with electrons on helium.

\subsection{Device description}
\label{sec:device}

\begin{figure}[h]
\includegraphics[width=0.65\linewidth]{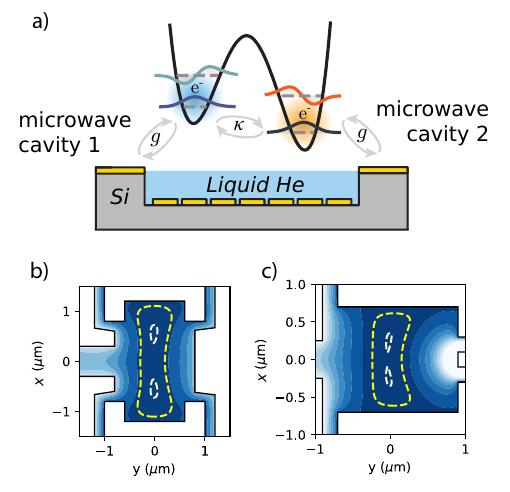}
\caption{Schematic layout of the electrons-on-helium device. Two electrons (e$^-$) are trapped on the surface of liquid helium in two adjacent electrostatic wells created by electrodes beneath the helium. Each electron is coupled ($g$) to its own microwave cavity (allowing single-qubit control and readout), and the two electrons interact via the Coulomb coupling $\kappa$ (enabling for example two-qubit gates \cite{leinonen2025}). Panels (b) and (c) show quantum dot geometries from experiments by Castoria et al.~\cite{castoria2025} and Koolstra et al.~\cite{koolstra2026strong}, respectively, illustrating electrostatic realizations of the double-well potential. Dashed lines indicate equipotentials.}
\label{fig:fig1}
\end{figure}

In Refs.~\cite{beysengulov2024,leinonen2025} we have studied a microdevice (see Fig.~\ref{fig:fig1} for a schematic layout) in which two electrons are trapped on the surface of liquid helium in an electrostatic double-well potential created by a set of micro-fabricated electrodes. Each electron, being near the helium surface, induces a slight polarization of the helium atoms, creating a weak attractive force toward the surface. However, at the helium–vacuum interface the electron encounters a large potential barrier (on the order of 1~eV) that prevents it from entering the liquid. As a result, the electron has a series of quantized bound states perpendicular to the surface (resembling a Rydberg-like spectrum)~\cite{monarkha2013two}. At cryogenic temperatures ($T<2~\mathrm{K}$), each electron remains in the ground state of this out-of-plane motion. The in-plane motion (parallel to the surface) is confined by the electrostatic potential of the trap. The device is designed so that the $x$-direction (along the line connecting the two wells) is the principal axis for in-plane motion, with level spacings in the range of 5--15~GHz. Motion in the orthogonal $y$-direction has a much higher energy scale (about six times larger frequency), so the $y$-motion can almost be neglected in the qubit operating space as a first approximation. Our approach described below deals however with the full two-dimensional complexity of the problem. Each electron is dipole-coupled to a separate microwave cavity for qubit readout and single-qubit control. The two electrons are separated by approximately $1.5~\mu\mathrm{m}$ in this device, and their Coulomb repulsion provides a controllable inter-qubit interaction, which we will exploit to generate entanglement, together with the spin degrees of freedom.

The dynamics of the two-electron system can be described by the two-dimensional  ($\hbar^2/2m_e =1$)
\begin{equation}
\hat{H} = \sum_{i=1}^{2} \Bigl( -\frac{1}{2}\,\nabla^2 + v(\bm{r}_i) \Bigr) + u(\bm{r}_1, \bm{r}_2)~,
\label{eq:H}
\end{equation}
which includes the kinetic energy and confining potential $v(\bm{r}_i)$ for each electron ($i=1,2$) (the one-body part) and the two-body interaction $u(\bm{r}_1,\bm{r}_2)$ between electrons. Here $u(\bm{r}_1,\bm{r}_2)$ represents the Coulomb repulsion between an electron at position $\bm{r}_1$ and another at $\bm{r}_2$. We have defined 
\begin{equation}
u(\bm{r}_1,\bm{r}_2) = \frac{\kappa}{\sqrt{\vert\bm{r}_1 - \bm{r}_2\vert^2 + \epsilon^2}},
\label{eq:coulombpot}
\end{equation}
where $\kappa$ sets the interaction strength (in appropriate units) and a small regularization parameter $\epsilon = 0.01$ is included to avoid potential  singularities at $\bm{r}_1 = \bm{r}_2$. 

The trapping potential $v(\bm{r})$ is created by various gate electrodes beneath the helium layer, see  Fig.~\ref{fig:fig1} for a schematic illustration. The experimental parameters that define the trapping potential are discussed below, in connection with the reinforcement-learning and optimization procedure. 
In Ref.~\cite{leinonen2025} we demonstrated how to prepare entangled states in order to construct two-qubit gates.
The double-well potential configurations can be physically realized in the electron-on-helium platform through electrostatic gating. By patterning electrodes beneath the superfluid helium surface, one can shape the electrostatic potential landscape experienced by a trapped electron, forming a tunable double-well profile analogous to those shown in Fig.~\ref{fig:fig1} (b) and (c). This approach has been demonstrated experimentally in recent works~\cite{castoria2025, koolstra2026strong}, where lithographically defined gate geometries produce the requisite potential barrier between two charge-stable electron trapping sites, though thus far only single-electron properties have been characterized.

\begin{figure}[htb!]
\centering
\includegraphics[width=\textwidth]{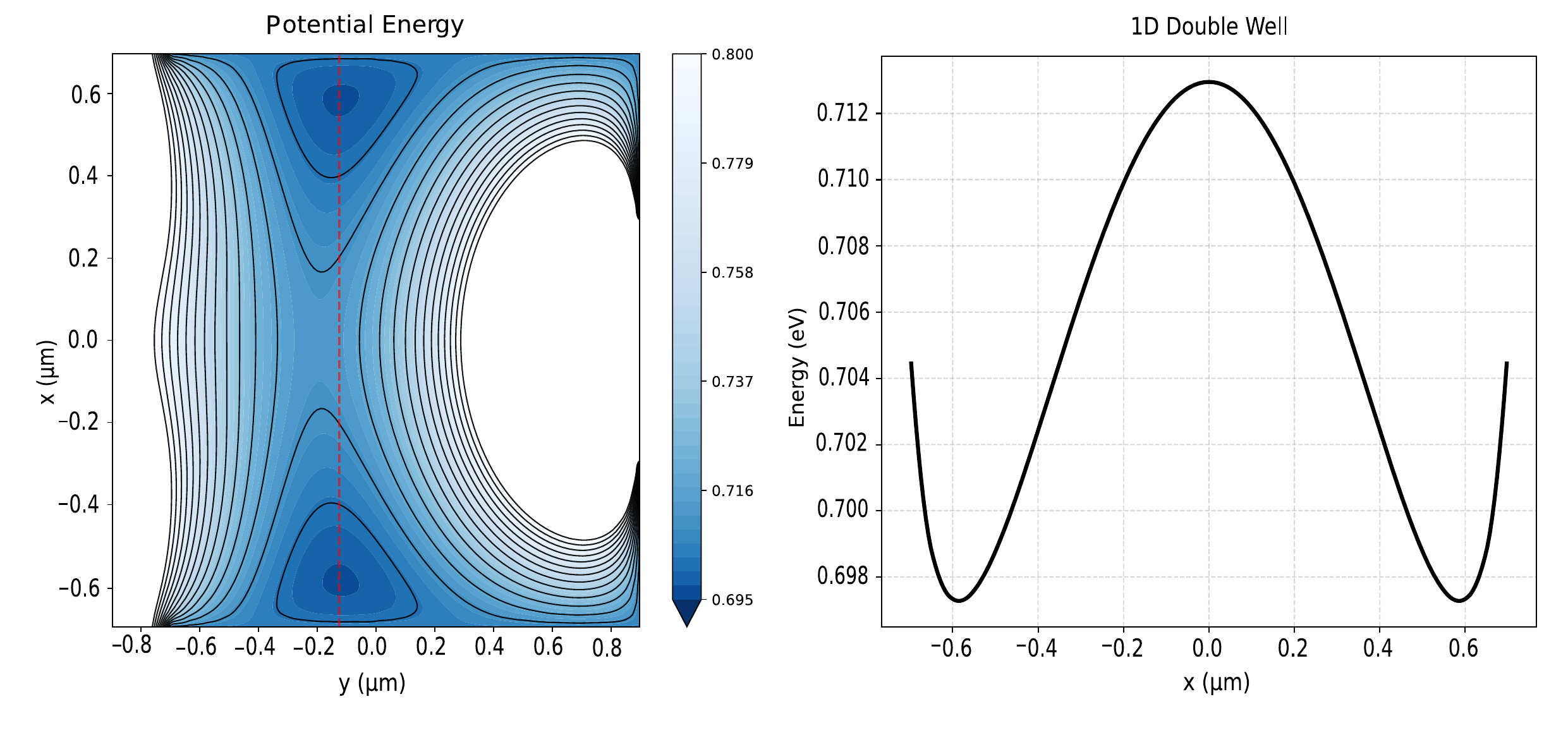}
\caption{Profile of the double-well obtained with the RL optimization technique having reward Eq. \eqref{eq:reward_function}. \emph{Left:} 2D representation of the potential energy (expressed in $\text{eV}$) . \emph{Right:} 1D energy profile of the double well (expressed in eV) extracted along the cut indicated by the red dashed line ($y \approx -0.125 \text{ \textmu m}$).}
\label{fig:double-well_RL}
\end{figure}

The optimized voltage configuration obtained with reinforcement learning (RL), see discussion below, generates a double-well structure shown in Fig.~\ref{fig:double-well_RL}. The 2D contour plot (left panel) displays the potential energy, while the 1D energy profile (right panel) highlights the double well along a cut taken at fixed $y \approx -0.125~\mu\mathrm{m}$ (the red dashed line in Fig.~\ref{fig:double-well_RL}).

The confinement induced by this double-well trap leads to the formation of localized single-particle orbitals,  as illustrated in Fig. \ref{fig:orbitals_RL}.

\subsection{Single-particle spectrum and localized orbitals}
\label{sec:spectrum}

We solve the single-particle problem on a discrete variable representation (DVR) grid \cite{Light1985}. The kinetic operator is represented in the sine-DVR, which on an $N$-point grid covering (here we use the $x$-axis as example, similar equations pertain to the $y$-axis) $[x_0,x_L]$, with points $x_j = x_0 + jL/(N+1)$ and weights $w_j = L/(N+1)$, yields the exact one-dimensional kinetic matrix
\begin{equation}
  T^{(\mathrm{1D})}_{jk} = \frac{1}{2}(-1)^{j-k}
  \begin{cases}
    \dfrac{\pi^2}{3}\!\left(\dfrac{N+1}{L}\right)^{\!2}
    - \dfrac{1}{L^{2}\sin^{2}[\pi j/(N+1)]}, & j=k,\\[10pt]
    \dfrac{1}{L^{2}}\!\left[\dfrac{1}{\sin^{2}[\pi(j-k)/(2(N+1))]}
         -\dfrac{1}{\sin^{2}[\pi(j+k)/(2(N+1))]}\right], & j\neq k,
  \end{cases}
  \label{eq:dvr_T}
\end{equation}
with $\hbar^{2}/2m=1$ in our units (consistent with Eq.~\eqref{eq:H}). The two-dimensional kinetic matrix is the Kronecker sum $T_{2\mathrm D} = T_x\otimes I_y + I_x\otimes T_y$, and the single-particle Hamiltonian is $H_{\rm sp} = T_{2\mathrm D} + V_{2\mathrm D}$, with $V_{2\mathrm D}$ representing the potential $v(\bm r)$ from Eq.~(\ref{eq:H}) on the grid. We use $N_x=N_y=20$ (400 grid points) and diagonalize the one-body problem in this basis in order to obtain all single-particle states. These states are subsequently used in the full configuration interaction theory calculation \cite{SzaboOstlund1996} discussed below.

The single-particle Hamiltonian $H_{\rm sp}$ is first diagonalized in the numerical basis, yielding eigenpairs $\{\varepsilon_k,\psi_k\}$.  
In the case of a nearly symmetric double well, which will be the main focus of our analysis, the eigenstates of $H_{\rm sp}$ are typically
symmetric and antisymmetric combinations spread over both wells, even though the physically relevant single-well orbitals are left- and right-localized combinations of these states.  We therefore define localized orbitals by diagonalizing a spatial projection operator within the chosen low-energy single-particle subspace.

\begin{figure}[htb!]
\centering
\includegraphics[width=\textwidth]{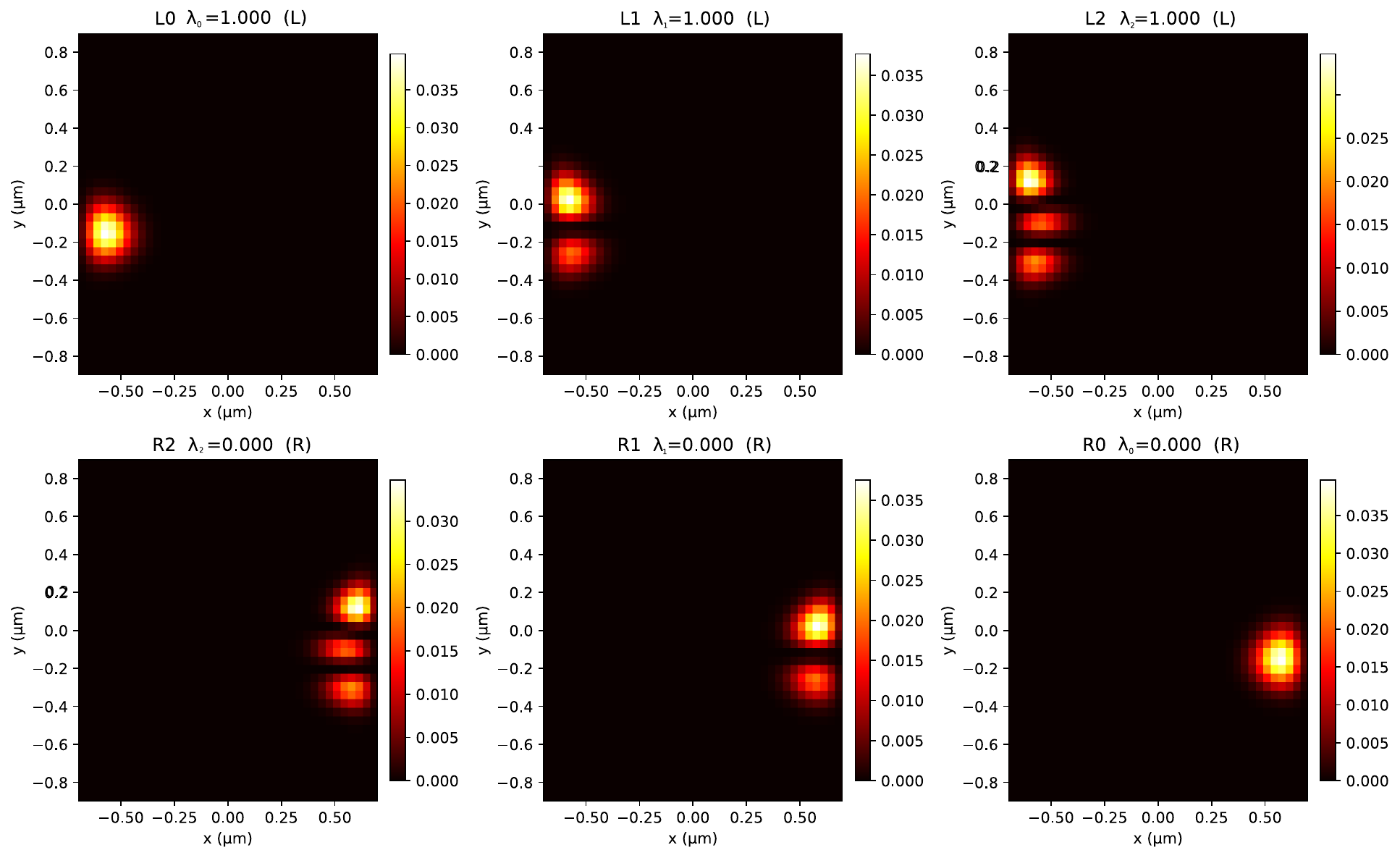}
\caption{Top: One-body probabilities for the first three leftmost localized orbitals L0,L1,L2 (left to right). Bottom: One-body
probabilities for the first three rightmost localized orbitals,R0, R1, and R2 (in reverse order).}
\label{fig:orbitals_RL} 
\end{figure}

Let $P_L$ denote the projector onto the left well,
\begin{equation}
  P_L = \int_{x<x_b} dx\, |x\rangle\langle x|,
\end{equation}
where $x_b$ is the position of the barrier separating the two wells, taken to be $x_b=0$ in the present geometry.  Equivalently, $P_R=1-P_L$ projects onto the right well.  In the truncated basis of single-particle eigenstates $\{\psi_k\}$ we form the matrix
\begin{equation}
  (P_L)_{ij}
  =
  \langle \psi_i | P_L | \psi_j\rangle
  =
  \int_{x<x_b} dx\, \psi_i^*(x)\psi_j(x).
\end{equation}
Diagonalizing this matrix gives states
\begin{equation}
  |\phi_\alpha\rangle
  =
  \sum_k U_{k\alpha} |\psi_k\rangle,
  \qquad
  P_L|\phi_\alpha\rangle
  =
  \lambda_\alpha |\phi_\alpha\rangle ,
\end{equation}
where $\lambda_\alpha\in[0,1]$ measures the weight of $|\phi_\alpha\rangle$ in the left well.  States with $\lambda_\alpha\simeq 1$ are labeled left-localized, states with $\lambda_\alpha\simeq 0$ are labeled right-localized, and states with intermediate $\lambda_\alpha$ are regarded as delocalized or ambiguous within the selected subspace.
The localized orbitals are then ranked by their single-particle energy expectation
values,
\begin{equation}
  \tilde{\varepsilon}_\alpha
  =
  \langle \phi_\alpha | H_{\rm sp} | \phi_\alpha\rangle ,
\end{equation}
giving the ordered labels $\mathrm{L0},\mathrm{L1},\dots$ (with $L$ representing the left well and $R$ representing the right well) and
$\mathrm{R0},\mathrm{R1},\dots$.  This projection-based construction is the one used below to classify the two-electron states.  It avoids the ambiguity that arises when the eigenstates of $H_{\rm sp}$ are delocalized bonding/antibonding combinations, while reducing to the usual left/right classification, based on the expectation value of the position operator $\braket{\hat{x}}$, when the single-particle eigenstates are already well localized. 
Figure \ref{fig:orbitals_RL} shows the various  one-body probabilities as functions of $x$ and $y$. The state $\mathrm{L0}$ is the lowest lying single-particle state (in energy) of the left well, while the state $\mathrm{R0}$ is the lowest-lying state for the right well. Introducing the two-body interaction described in Eq.~(\ref{eq:coulombpot}) and diagonalizing the two-body problem in this single-particle basis, will give rise to two-body states with electrons localized in each well. If we assign the states $\mathrm{R0}$ and $\mathrm{L0}$ to be in a qubit $0$ state, the lowest-lying two-body state could then be a singlet state with the qubit assignment $\vert 0\rangle_L \otimes \vert 0\rangle_R=\vert 00\rangle$. In a similar vein, the states $\mathrm{L1}$ and $\mathrm{R1}$ can be used to define a two-body state with qubit $1$, that is $\vert 1\rangle_L \otimes \vert 1\rangle_R=\vert 11\rangle$.

\subsection{Two-electron states and interaction-generated entanglement}
\label{sec:cientangle}

The interacting two-electron problem is solved by configuration interaction (CI) \cite{SzaboOstlund1996}.
Starting from the low-energy eigenstates of the single-particle Hamiltonian $H_{\rm sp}$, we construct a spin-adapted two-electron basis containing singlet and triplet configurations.

For two spin-$\tfrac12$ electrons occupying spatial orbitals $a$ and $b$ (the above states $\mathrm{L0},\mathrm{L1},\dots$ and $\mathrm{R0},\mathrm{R1},\dots$), the Pauli principle requires the total wave function to be antisymmetric under particle exchange. Thus a spatially symmetric orbital wave function must be combined with the antisymmetric spin singlet,
\begin{equation}
  \ket{\chi_S}
  =
  \frac{1}{\sqrt{2}}
  \left(
  |\uparrow\downarrow\rangle
  -
  |\downarrow\uparrow\rangle
  \right),
\end{equation}
whereas a spatially antisymmetric orbital wave function is combined with one of the three symmetric spin triplets,
\begin{align}
  \ket{\chi_{T_+}} &= |\uparrow\uparrow\rangle,\\
  \ket{\chi_{T_0}} &=
  \frac{1}{\sqrt{2}}
  \left(
  |\uparrow\downarrow\rangle
  +
  |\downarrow\uparrow\rangle
  \right),\\
  \ket{\chi_{T_-}} &= |\downarrow\downarrow\rangle .
\end{align}
For two distinct orbitals $a\neq b$, the corresponding spatial factors are
\begin{align}
  \ket{\Phi^{(+)}_{ab}(1,2)}
  &=
  \frac{1}{\sqrt{2}}
  \left[
  \ket{\psi_a(1)\psi_b(2)}
  +
  \ket{\psi_b(1)\psi_a(2)}
  \right],
  \\
  \ket{\Phi^{(-)}_{ab}(1,2)}
  &=
  \frac{1}{\sqrt{2}}
  \left[
  \ket{\psi_a(1)\psi_b(2)}
  -
  \ket{\psi_b(1)\psi_a(2)}
  \right].
\end{align}
The singlet basis states are therefore
$\ket{\Phi^{(+)}_{ab}\chi_S}$, while the triplet basis states are
$\ket{\Phi^{(-)}_{ab}\chi_{T_m}}$ with $m=+,0,-$.  If both electrons occupy the same spatial orbital, $a=b$, the antisymmetric spatial combination vanishes, so only the singlet configuration is allowed. In the absence of spin-dependent interactions the three triplet spin projections are degenerate, and it is sufficient to diagonalize one triplet block together with the singlet block. The basis that will be used in the diagonalization of the two-electron problem is truncated by ordering two-particle configurations according to the one-body energy sum $\varepsilon_a+\varepsilon_b$ and retaining the lowest $N_{\rm CI}$ configurations.  This energy-based truncation avoids excluding low-energy mixed-orbital configurations that may be important near avoided crossings.

The two-electron Hamiltonian is defined in Eq.~\eqref{eq:H}, where the electron--electron interaction is represented on the numerical grid by the softened Coulomb interaction defined in Eq.~\eqref{eq:coulombpot}.
Diagonalizing the CI Hamiltonian $H$ gives eigenstates $|\Psi_n\rangle=\sum_I C^n_{I}|I\rangle$, where $|I\rangle$ denotes a spin-adapted two-electron configuration.

The analysis of the CI eigenvectors is performed in terms of the projection-localized orbitals defined above.  This allows each eigenstate to be resolved into left--left, left--right, and right--right charge sectors, as well as
into orbital components such as $\mathrm{Lm}$, $\mathrm{Rn}$, where  $\mathrm{L}$, ($\mathrm{R}$) denote the left- (right-) character of the orbital, and  $m=0,1,\dots$, $n=0,1,\dots$ its local (spatial) index. 

The CI eigenvectors are analyzed in the projection-localized single-particle basis defined in Sec.~\ref{sec:spectrum}.  We use the labels $\mathrm{L0},\mathrm{L1},\dots$ and $\mathrm{R0},\mathrm{R1},\dots$ introduced there to resolve the orbital content of each two-electron eigenstate.  We note that the
singlet-like and triplet-like states used for sensing need not be described by the
bare $\mathrm{L0}$ and $\mathrm{R0}$ orbitals alone, since in the full CI calculation the
dominant left and right spatial modes can be coherent superpositions of several
projection-localized orbitals.
For this reason, when comparing a CI eigenstate with an idealized singlet or triplet, we do not choose the left and right orbitals a priori to be simply $\mathrm{L0}$ and $\mathrm{R0}$.  Instead, we let the many-body state itself define the optimal left and right spatial modes. This is done through the spin-summed one-body density matrix.  The idea is that the one-body density matrix contains the complete single-particle information of the correlated two-electron state: its diagonal elements give the occupation of the localized orbitals, while its off-diagonal elements encode coherent mixing between them. After rotating the CI eigenstate to the projection-localized basis, we
construct the spin-orbital one-body density matrix
\begin{equation}
  \rho^{(1)}_{pq}
  =
  \bra{\Psi}
  c_q^\dagger c_p
  \ket{\Psi},
\end{equation}
where $p,q$ label localized spin-orbitals.  We then trace over spin to obtain a
spatial one-body density matrix,
\begin{equation}
  \rho^{(1)}_{\mu\nu}
  =
  \rho^{(1)}_{\mu\uparrow,\nu\uparrow}
  +
  \rho^{(1)}_{\mu\downarrow,\nu\downarrow},
\end{equation}
where $\mu,\nu$ label projection-localized spatial orbitals.  This object tells us which spatial orbitals are actually occupied by the correlated CI state, independently of the spin configuration.

We then restrict $\rho^{(1)}_{\mu\nu}$ to the left and right subspaces separately.
That is, we form the blocks
\begin{equation}
  \rho^{(1)}_{LL}
  =
  \left.
  \rho^{(1)}
  \right|_{\mu,\nu\in L},
  \qquad
  \rho^{(1)}_{RR}
  =
  \left.
  \rho^{(1)}
  \right|_{\mu,\nu\in R}.
\end{equation}
Diagonalizing these two blocks gives the natural orbitals within the left and right wells.  The eigenvector with the largest eigenvalue in the left block is the normalized left-localized orbital with the largest occupation in the CI state; the corresponding eigenvector in the right block gives the dominant right-localized orbital.  In this sense, the dominant natural orbitals are the best single-orbital left/right approximation to the correlated two-electron state. Diagonalizing the left and right blocks gives the dominant natural orbitals
\begin{equation}
  \ket{\bar\Phi_L}
  =
  \sum_{\mu\in L}
  u^{(L)}_\mu \ket{\phi_\mu},
  \qquad
  \ket{\bar\Phi_R}
  =
  \sum_{\nu\in R}
  u^{(R)}_\nu \ket{\phi_\nu},
\end{equation}
where $u^{(L)}$ and $u^{(R)}$ are the eigenvectors with the largest occupations in the left and right blocks, respectively.  These orbitals define the best single-orbital left/right description of the correlated CI state.

Using these dominant natural orbitals, the reference singlet and triplet states are
\begin{align}
  \ket{S^\ast}
  &=
  \frac{1}{\sqrt{2}}
  \left(
  d^\dagger_{L\uparrow}d^\dagger_{R\downarrow}
  -
  d^\dagger_{L\downarrow}d^\dagger_{R\uparrow}
  \right)\ket{0},
  \\
  \ket{T^\ast_0}
  &=
  \frac{1}{\sqrt{2}}
  \left(
  d^\dagger_{L\uparrow}d^\dagger_{R\downarrow}
  +
  d^\dagger_{L\downarrow}d^\dagger_{R\uparrow}
  \right)\ket{0},
\end{align}
with
\begin{equation}
  d^\dagger_{L\sigma}
  =
  \sum_{\mu\in L}
  u^{(L)}_\mu c^\dagger_{\mu\sigma},
  \qquad
  d^\dagger_{R\sigma}
  =
  \sum_{\nu\in R}
  u^{(R)}_\nu c^\dagger_{\nu\sigma}.
\end{equation}
The overlaps of the CI eigenstates with $\ket{S^\ast}$ and $\ket{T^\ast_0}$ then measure how close the many-body states are to ideal left--right singlet and triplet configurations.

The projection-localized orbitals define the left and right subspaces, while the dominant natural orbitals identify the actual spatial modes occupied by the correlated CI state.  The ideal operating regime is therefore one in which the relevant singlet-like and triplet-like CI eigenstates have large overlap with $\ket{S^\ast}$ and $\ket{T^\ast_0}$, respectively, and only weak residual weight in configurations not captured by these dominant left/right modes.

\subsection{Signal encoding by a magnetic-field gradient}
\label{sec:signal_encoding}

In order to demonstrate the capability of electrons on helium as potential quantum sensing devices, we limit ourselves in this work to the detection of a static magnetic-field gradient across the two wells. Since the relevant two-electron states are predominantly in the $\mathrm{L\textrm{m}}-\mathrm{R\textrm{n}}$ charge sector (prepared as a spin-singlet state), the gradient is sensed through the different Zeeman fields experienced by the left- and right-localized spins.  The magnetic
coupling is
\begin{equation}
  H_B
  =
  g\mu_B
  \left(
  B_L S^z_L + B_R S^z_R
  \right),
  \label{eq:HB_local}
\end{equation}
where $B_L$ and $B_R$ are the magnetic fields at the left and right wells.  Writing $\bar B = \left(B_L+B_R\right)/2$, $ \delta B = B_L-B_R$,
we obtain
\begin{equation}
  H_B
  =
  g\mu_B \bar B
  \left(S^z_L+S^z_R\right)
  +
  \frac{g\mu_B\delta B}{2}
  \left(S^z_L-S^z_R\right).
  \label{eq:HB_uniform_gradient}
\end{equation}
The uniform field term shifts the triplet manifold according to its total spin projection, but it does not mix the singlet and the $m=0$ triplet.  By contrast, the gradient term is antisymmetric under interchange of the two dots and couples the singlet $S_0^*$ to the $T^*_0$ triplet.

In the idealized $\mathrm{L0}\mathrm{R0}$ spin subspace, the relevant states are
$
  \ket{S} =
  \frac{1}{\sqrt{2}}
  \left(
  \ket{\uparrow_L\downarrow_R}
  -
  \ket{\downarrow_L\uparrow_R}
  \right)
  $ and $
  \ket{T_0}=
  \frac{1}{\sqrt{2}}
  \left(
\ket{\uparrow_L\downarrow_R}
  +
 \ket{\downarrow_L\uparrow_R}
  \right).$
Furthermore, the gradient operator satisfies
$ \left(S^z_L-S^z_R\right)\ket{S} = \ket{T_0}$ ,
$ \left(S^z_L-S^z_R\right)\ket{T_0} = \ket{S}$,
so that
\begin{equation}
  \langle T_0|H_B|S\rangle
  =
  \frac{g\mu_B\delta B}{2}.
  \label{eq:ST_gradient_coupling}
\end{equation}
Thus the magnetic-field gradient directly drives coherent mixing between the
singlet-like ground state and the triplet-like excited state.

In the full CI calculation the relevant states are not perfectly pure
$\mathrm{L0}\mathrm{R0}$ singlet and triplet configurations.  We therefore project
the Zeeman operator onto the two CI eigenstates of interest: the ground state
$|g\rangle$, which is predominantly singlet-like, and an excited state $|e\rangle$,
which is predominantly triplet-like.  The effective two-level Hamiltonian is
\begin{equation}
  H_{\rm eff}
  =
  E_g |g\rangle\langle g|
  +
  E_e |e\rangle\langle e|
  +
  \lambda(\delta B)
  \left(
  |g\rangle\langle e| + |e\rangle\langle g|
  \right),
  \label{eq:effective_ST_hamiltonian}
\end{equation}
with
\begin{equation}
  \lambda(\delta B)
  =
  \frac{g\mu_B\delta B}{2}
  \,
  \mathcal M_{ge},
  \qquad
  \mathcal M_{ge}
  =
  \langle e|
  S^z_L-S^z_R
  |g\rangle .
  \label{eq:gradient_matrix_element}
\end{equation}
For an ideal, isolated $\mathrm{L0}-\mathrm{R0}$ singlet--triplet pair,
$\mathcal M_{ge}=1$.  In the numerical calculation, deviations of
$\mathcal M_{ge}$ from unity quantify the reduction caused by charge admixture,
orbital excitation, or imperfect localization.

If the system is prepared in the singlet-like ground state, the field gradient is
encoded in the subsequent coherent oscillation into the triplet-like state.  Defining
the singlet--triplet detuning
\begin{equation}
  \Delta_{ST}=E_e-E_g ,
\end{equation}
the transition probability is, within the two-level approximation,
\begin{equation}
  P_{g\rightarrow e}(t)
  =
  \frac{4\lambda^2}
  {\Delta_{ST}^2+4\lambda^2}
  \sin^2
  \left[
  \frac{t}{2\hbar}
  \sqrt{\Delta_{ST}^2+4\lambda^2}
  \right].
  \label{eq:gradient_transition_probability}
\end{equation}
The magnetic gradient can therefore be estimated either from the transition
probability after a fixed interrogation time or, equivalently, from the coherent
singlet--triplet oscillation frequency.  This identifies $\delta B$ as the unknown
parameter encoded in the quantum state by the Zeeman-gradient Hamiltonian, making
the problem one of quantum-parameter estimation.  The
comparison with two independent spin probes is made below, after introducing the
quantum Fisher information and the corresponding optimal-measurement framework.

\subsection{Parameter estimation and optimal measurements}
\label{sec:qfi-est}
We now quantify the sensitivity of the magnetic-gradient protocol introduced in
Sec.~\ref{sec:signal_encoding}.  We first recall the general parameter-estimation
framework~\cite{wiseman_quantum_2010}.  Let $\varphi$ be an unknown parameter encoded in a quantum state
$\ket{\psi(\varphi)}$.  For an unbiased estimator $\varphi_{\rm est}$ obtained from
$\nu$ independent repetitions, the quantum Cramér--Rao bound gives
\begin{equation}
  \mathrm{Var}(\varphi_{\rm est})
  \geq
  \frac{1}{\nu F_Q(\varphi)} .
  \label{eq:qcrb_general}
\end{equation}
Here $F_Q(\varphi)$ is the quantum Fisher information~\cite{braunstein1994statistical,paris2009quantum}. For a pure state, the quantum Fisher information is
\begin{equation}
  F_Q(\varphi)
  =
  4
  \left[
  \braket{\partial_\varphi\psi|\partial_\varphi\psi}
  -
  \left|
  \braket{\psi|\partial_\varphi\psi}
  \right|^2
  \right],
  \label{eq:qfi_pure_general}
\end{equation}
where $\ket{\partial_\varphi\psi}=\partial_\varphi\ket{\psi(\varphi)}$.  For a
specific measurement with outcome probabilities $p_m(\varphi)$, the corresponding
classical Fisher information is
\begin{equation}
  F_C(\varphi)
  =
  \sum_m
  \frac{\left[\partial_\varphi p_m(\varphi)\right]^2}{p_m(\varphi)} ,
  \label{eq:classical_fisher_general}
\end{equation}
with $F_C(\varphi)\leq F_Q(\varphi)$.  The optimal measurement is the one for which
this bound is saturated.

We now specialize the general parameter-estimation framework to the magnetic-gradient
protocol by setting $\varphi\equiv\delta B$. 
As described in Sec.~\ref{sec:signal_encoding}, the gradient couples the
singlet-like ground state $\ket{g} \equiv \ket{S^\ast}$ to the triplet-like excited state $\ket{e}\equiv \ket{T^\ast_0}$,
with coupling matrix element $\mathcal M_{ge}$ and splitting
$\Delta_{ST}=E_e-E_g$.  A natural readout is therefore the transition probability
$P_{g\to e}(\delta B,t)$, i.e. the triplet-like population after preparing the
system in $\ket{g}$.  The corresponding binary measurement
$\{\ket{e}\bra{e},\mathbb{1}-\ket{e}\bra{e}\}$ has classical Fisher information
\begin{equation}
  F_C
  =
  \frac{
  \left(\partial_{\delta B}P_{g\to e}\right)^2
  }{
  P_{g\to e}\left(1-P_{g\to e}\right)
  }
  \label{eq:binary_fisher}
\end{equation}
with $P_{g\to e}$ computed from the two-level Rabi expression
\begin{equation}
  P_{g\to e}(\delta B,t)
  =
  \frac{4\lambda^2}{\Delta_{ST}^2+4\lambda^2}
  \sin^2\!\left[
  \frac{t}{2\hbar}
  \sqrt{\Delta_{ST}^2+4\lambda^2}
  \right],
  \qquad
  \lambda(\delta B)
  =
  \frac{g\mu_B\delta B}{2}\mathcal M_{ge}.
  \label{eq:Pge_rabi_full}
\end{equation}
$F_C(\varphi)$ retains the oscillatory structure of the readout sensitivity.
In particular, it develops dips at interrogation
times for which the measured population is locally insensitive to the parameter,
i.e. when $\partial_{\delta B}P_{g\to e}$ vanishes.  These dips are a characteristic feature of the  singlet--triplet readout and can be seen in the
numerical evaluation of $F_C$ (see Figs.~\ref{fig:populations}-\ref{fig:RamseyEntDiff} below).

Conversely, $F_Q$ can be evaluated from Eq.~\eqref{eq:qfi_pure_general}, using the effective dynamics of Sec.~\ref{sec:signal_encoding},
\begin{equation}
  F_Q^{ST}(t)
  =
  \frac{4(g\mu_B)^2|\mathcal M_{ge}|^2}{\Delta_{ST}^2}
  \sin^2\left(\frac{\Delta_{ST}t}{2\hbar}\right),
  \label{eq:FQ_ST_detuned}
\end{equation}
which, for $\Delta_{ST}t/\hbar\ll1$, reduces to $
  F_Q^{ST}(t)
  \simeq
  |\mathcal M_{ge}|^2
  \left(\frac{g\mu_B t}{\hbar}\right)^2$ .

Thus the CI calculation enters the metrological figure of merit through the same
quantities identified in the signal-encoding analysis: the singlet--triplet matrix
element, the singlet--triplet splitting, and leakage outside the target
$\mathrm{L0}\mathrm{R0}$ sector.

\subsection{Independent-spin benchmark}

We compare the singlet--triplet protocol with a separable reference strategy in
which the two electrons are used as independent spin probes.  The same parameter
$\delta B=B_L-B_R$ is estimated, but now through ordinary Ramsey-type phase
accumulation rather than through the transition probability $P_{g\to e}$.

With the convention used above, the generator of translations in $\delta B$ is
\begin{equation}
  G_{\delta B}
  =
  \frac{g\mu_B t}{2\hbar}
  \left(S^z_L-S^z_R\right).
\end{equation}
For a pure state undergoing unitary parameter encoding, $F_Q=4\,\mathrm{Var}(G_{\delta B})$.
The optimal separable preparation is a product of equatorial spin states, e.g.
$(\ket{\uparrow}+\ket{\downarrow})/\sqrt{2}$ for each spin,  for which $ \mathrm{Var}\left(S^z_L-S^z_R\right)=1/2$.
Thus the independent-spin benchmark is
\begin{equation}
  F_Q^{\rm ind}(t)
  =
  \frac{1}{2}
  \left(
  \frac{g\mu_B t}{\hbar}
  \right)^2 .
  \label{eq:FQ_independent}
\end{equation}
This should be compared with the short-time, weak-gradient limit of the
singlet--triplet result derived above,
\begin{equation}
  F_Q^{ST}(t)
  \simeq
  |\mathcal M_{ge}|^2
  \left(
  \frac{g\mu_B t}{\hbar}
  \right)^2 .
  \label{eq:FQ_ST_compare}
\end{equation}
For an ideal isolated $\mathrm{L0}\mathrm{R0}$ singlet--triplet pair,
$\mathcal M_{ge}=1$, giving
$ F_Q^{ST}=2F_Q^{\rm ind}$.
The ideal entangled protocol therefore improves the Fisher information by a factor
of two.  In the full CI calculation this enhancement is reduced by
$|\mathcal M_{ge}|^2$ and by finite-detuning effects through
Eq.~\eqref{eq:FQ_ST_detuned}.

\subsection{Optimizing the entangled probe}
\label{sec:rl}

%\MEP{added some ref.s}

The double-well optimization problem can be mapped onto a Reinforcement Learning (RL) task, in which the training process, driven by agent-environment interaction, aims to maximize a reward function. Here we consider the electrostatic potential in a quantum dot geometry, realized in experiments by Koolstra et al.~\cite{koolstra2026strong}. There are four electrodes, controlling the potential and the physical parameters involved in this optimization are the electrode voltages $V=\{V_\mathrm{unload}, V_\mathrm{trap}, V_\mathrm{res}, V_\mathrm{barrier}\}$ that determine the double-well shape. Among the RL-based optimization algorithms used for quantum metrology \cite{bukov2026reinforcement}, we draw inspiration from  Belliardo \textit{et al.}~\cite{belliardo2024model}, and we implement an RL architecture based on the model-free Actor-Critic framework, consisting of an agent network and a critic network.  The former outputs a continuous action vector representing the electrode gate voltages for the trap and receives a reward based on the resulting trap shape after interacting with the environment and the latter evaluates the expected reward for the learning process. The episode immediately terminates and is reset for the next independent configuration. Specifically, we use the Proximal Policy Optimization (PPO) algorithm \cite{schulman2017proximal}, which ensures stable learning by constraining the size of parameter updates. Since the potential shape is highly sensitive to electrode voltage variations, the adopted strategy prevents large changes in the parameters, allowing for stable convergence. We formulate the task as a one-shot optimization problem, in which each episode is realized through a single interaction between the agent and the environment, starting from a parameter configuration randomly generated within the parameter range and then fed into both networks as the initial state.
In RL tasks, the reward function must directly reflect the optimization goal, which, in this case, is to maximize the Fisher Information (FI) during a sensing process. We write the reward as a function of the Two-Channel (Binary) Classical FI (CFI) and the Three-Channel (Full Spin-Resolved) CFI. During the agent-environment interaction, non-localized orbitals are identified and computed, along with the height of the central barrier, thereby introducing a physical constraint on the trap shape. Both terms are included as penalties in the reward function, constraining the RL to find solutions that preserve the double-well shape of the trap and reduce the number of delocalized or ambiguous states to zero. To compute the final reward, the evolution of the many-body states under a magnetic field perturbation ($\delta B= B_\mathrm{L} - B_\mathrm{R}$) is computed, and the maximum Classical FI for both channels is calculated. Note that we compute the total CFI for both configurations as the average of the CFIs obtained from the time evolution of two distinct initial many-body states. To maximize the sensitivity to the magnetic field perturbation, we also consider in the reward function the fidelities of the ground state with a singlet ($ F_{\mathrm{singlet}}$) and the first excited state with a triplet ($F_{\mathrm{triplet}}$). We build the final reward function related to the action vector $\mathbf{a}$  as follows
\begin{equation}
    R(\mathbf{a}) = r_\mathrm{f} \cdot r_\mathrm{cfi} - ( \omega_\mu \cdot p_\mu + \omega_\mathrm{b} \cdot p_\mathrm{b}),
    \label{eq:reward_function}
\end{equation}
where $r_\mathrm{f} = F_\mathrm{singlet} \cdot F_\mathrm{triplet}$ takes into account the fidelities, $r_\mathrm{cfi} = \mathrm{cfi}_3 + \alpha \cdot \mathrm{cfi}_{2}$ is related to the CFIs and $p_\mu$ and $p_\mathrm{b}$ respectively rely on the two weighted penalties, with weights $\omega_\mu$ and $\omega_\mathrm{b}$. The structure of the reward implemented in Eq. \eqref{eq:reward_function} prevents the RL agent from maximizing one metric at the expense of the other.
Table \ref{tab:ppo_hyperparameters} lists the hyperparameters used in our RL calculations.
\begin{table}[H]
\centering
\caption{Hyperparameters of the PPO agent used for the double-well potential optimisation.}
\label{tab:ppo_hyperparameters}
\begin{tabular}{llc}
\hline
\textbf{Hyperparameter} & \textbf{Description} & \textbf{Value} \\ \hline
Actor Optimizer & Algorithm for updating actor network weights & Adam ($3 \times 10^{-4}$) \\
Critic Optimizer & Algorithm for updating critic network weights & Adam ($1 \times 10^{-4}$) \\
Discount Factor ($\gamma$) & Temporal discount for future rewards (One-Shot) & 0.0 \\
GAE Parameter ($\lambda$) & Generalized Advantage Estimation parameter & 0.95 \\
Clip Range ($\epsilon$) & PPO policy objective clipping parameter & 0.2 \\
Entropy Coefficient & Weight of the entropy bonus in the loss function & 0.04 \\
Critic Loss Weight & Scaler for the value function loss component & 0.5 \\
Max Gradient Norm & Threshold for gradient clipping & 0.5 \\
Rollout Steps & Number of steps collected per environment per iteration & 512 \\
Parallel Environments & Number of environments running in parallel & 1 \\
Update Epochs & Number of optimization epochs per iteration & 8 \\
Minibatch Size & Size of the dataset subsets used for gradient steps & 64 \\ \hline
\end{tabular}
\end{table}

\subsection{Quantum and classical Fisher information with entangled electrons} 

\begin{figure}[htb!]
\centering
\includegraphics[width=\textwidth]{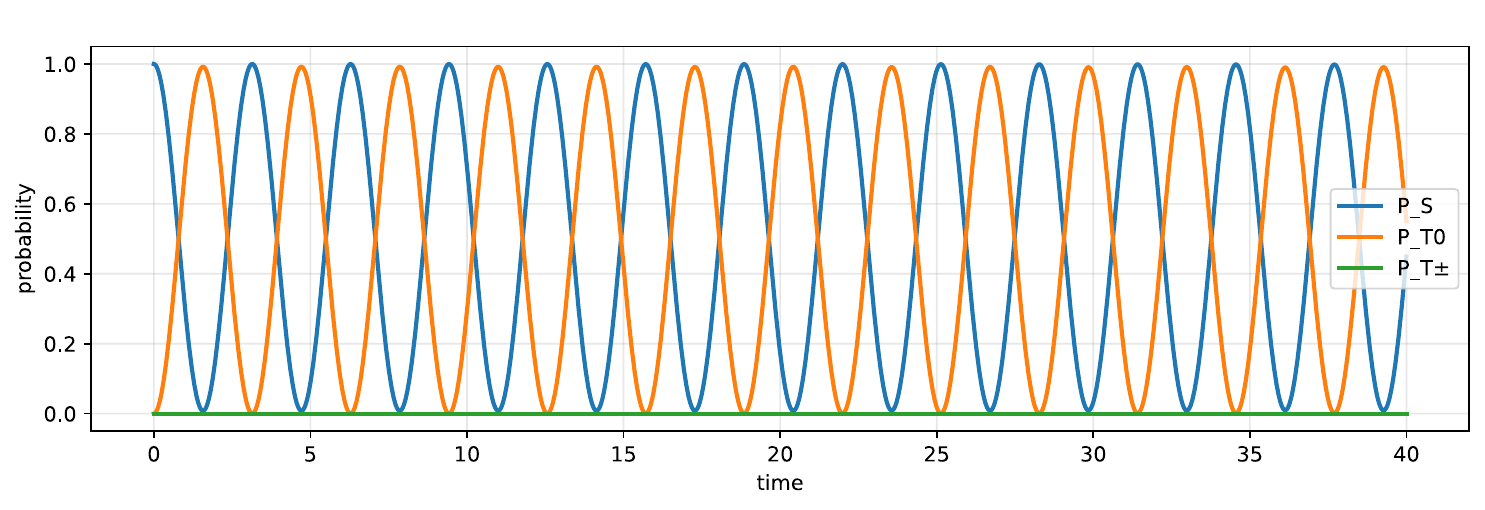}
\caption{Occupation probability of states $\ket{S^*}$, $\ket{T_0^*}$ and $\ket{T_\pm^*}$. The oscillations are induced by the coupling with the magnetic field ($B=1~\mu\mathrm{m}^{-2}$). For the perturbation parameters chosen and for the voltage configuration determined by the RL algorithm, it is possible to see that the full dynamics is well approximated by the effective two-level Hamiltonian~\eqref{eq:effective_ST_hamiltonian}.
Throughout the manuscript, unless otherwise stated, we use natural units with $\hbar^2/2m_e =1$ and $\hbar=1$, so that time is expressed in units of $\mu\mathrm{m}^2$.}
\label{fig:populations}
\end{figure}

\begin{figure}[htb!]
\centering
\includegraphics[width=0.9\textwidth]{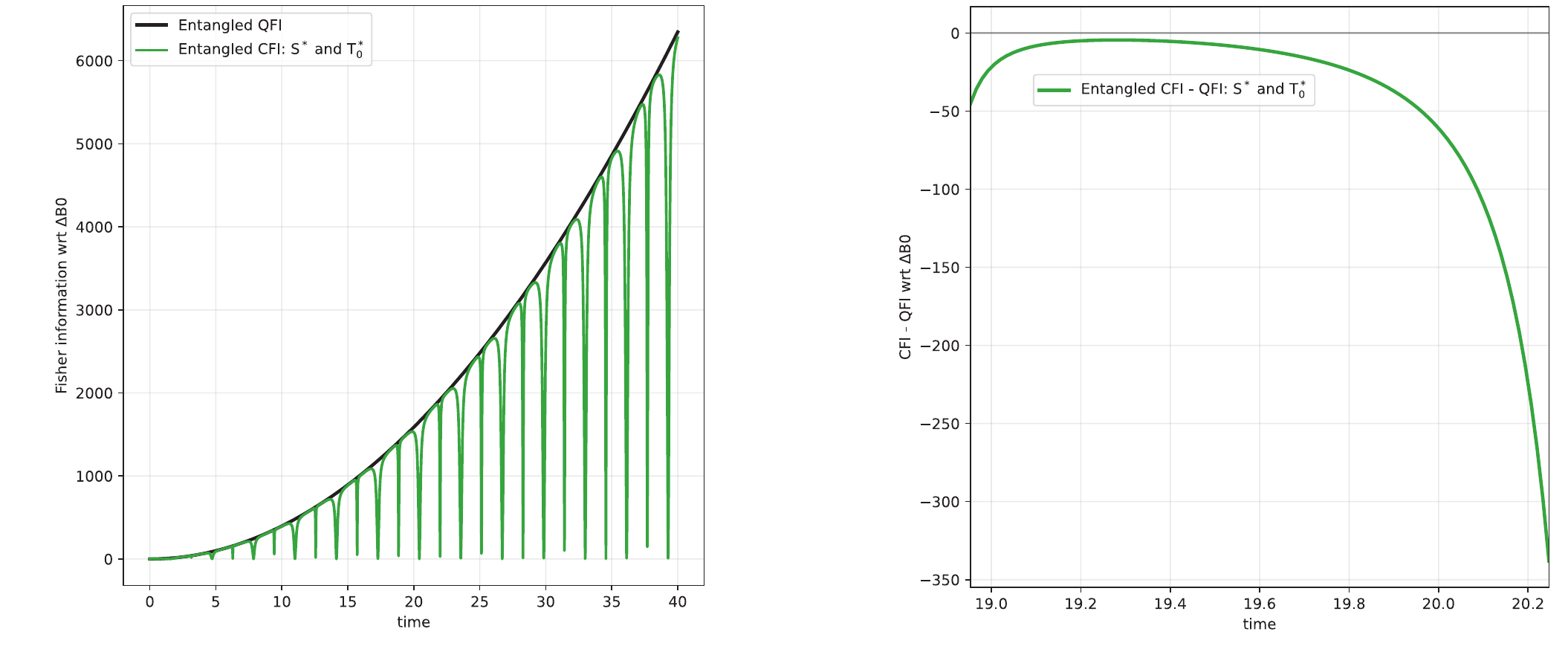}
\caption{\emph{Left:} Quantum and classical Fisher information for the entangled probe. Away from regions for which $\partial_{\delta B}P_{g \to e} \simeq 0$, the classical Fisher information saturates the quantum limit. \emph{Right:} Difference between quantum and classical Fisher information for the same probe.}
\label{fig:Ent-FI}
\end{figure}

\begin{figure}[htb!]
\centering
\includegraphics[width=0.9\textwidth]{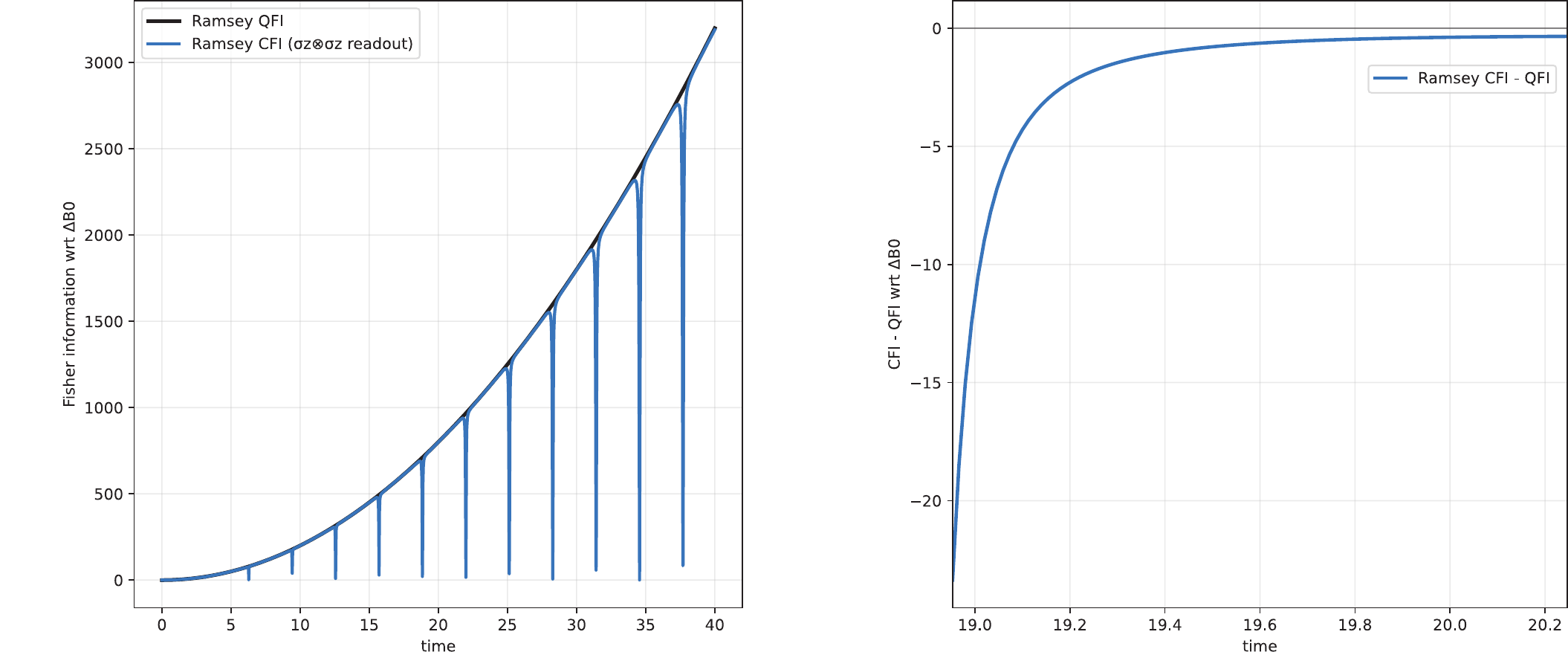}
\caption{Same as Fig.~\ref{fig:Ent-FI}, but now for the Ramsey setup. Note that, in this case, the Fisher information has a value which is half the value for the entangled probes case.}
\label{fig:Ramsey-FI}
\end{figure}

\begin{figure}[htb!]
\centering
\includegraphics[width=0.9\textwidth]{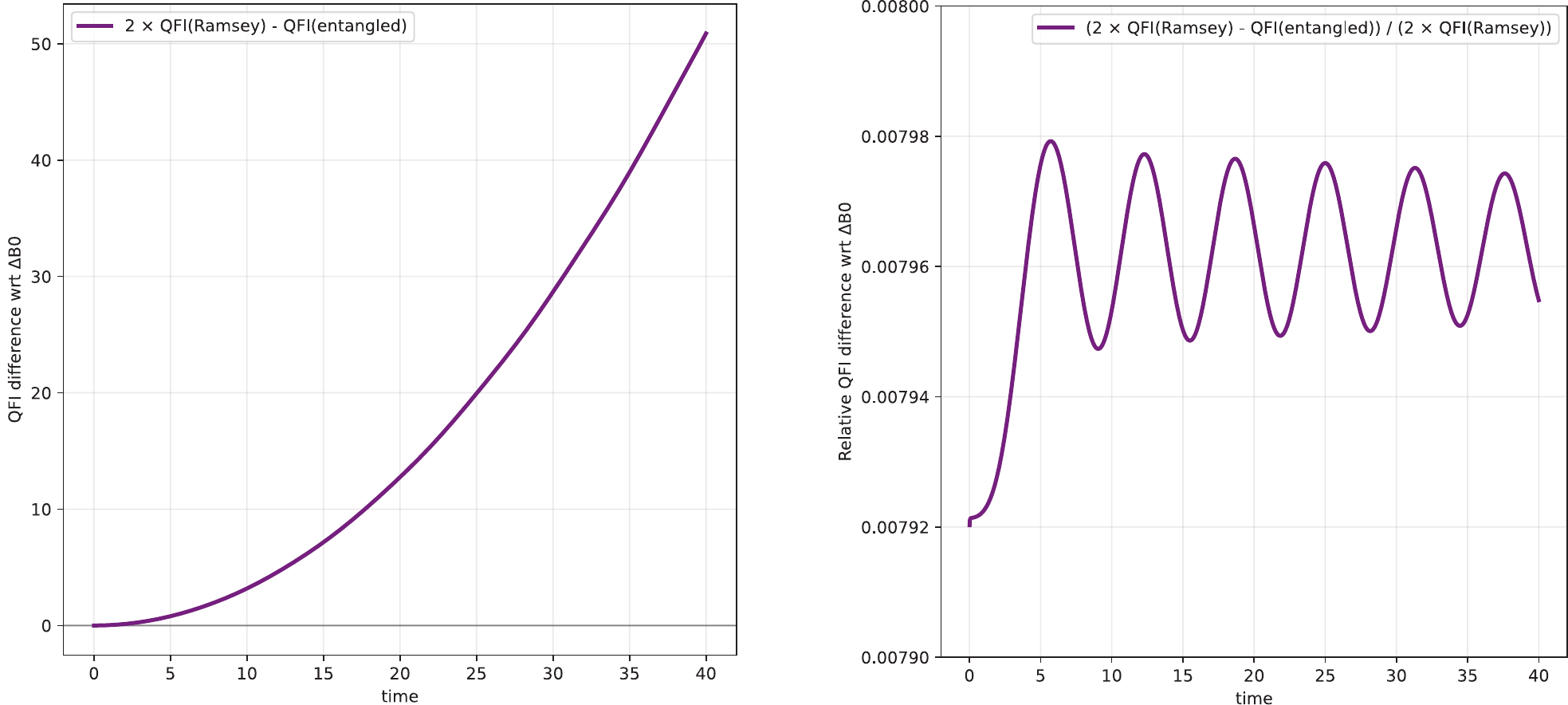}
\caption{Absolute (\emph{left}) and relative (\emph{right}) difference between the QFI of the Ramsey setup ($\times$ 2) and the entangled probes setup.}
\label{fig:RamseyEntDiff}
\end{figure}

In Figs.~\ref{fig:populations}-\ref{fig:RamseyEntDiff}, we summarize the results of the magnetic-gradient
estimation analysis.
In particular, we plot the quantum and classical Fisher information for
the correlated two-electron double-well protocol (Fig.~\ref{fig:Ent-FI}) and  the corresponding quantities
for the independent local Ramsey benchmark (Fig.~\ref{fig:Ramsey-FI}).

As expected, the classical Fisher information remains bounded by the quantum Fisher
information.  It also exhibits pronounced oscillations, including dips at
interrogation times where the measured transition probability is locally
insensitive to the magnetic-field gradient (see Eqs.~\eqref{eq:binary_fisher} and \eqref{eq:Pge_rabi_full} and Fig.~\ref{fig:populations}).
In the optimal regime, the quantum Fisher information of the correlated singlet--triplet 
protocol approaches twice that of the independent Ramsey benchmark of Fig.~\ref{fig:Ramsey-FI}, with a relative difference of less than 1\%, corresponding to the expected gain in sensitivity.
 The above results demonstrate clearly that for a time-independent field, entanglement due to spin and motional degrees of freedom yield an expected increase in sensitivity. 
 We will address time-dependent fields in  future studies.

\subsection{From two-qubit sensors to correlated sensor networks}
\label{sec:network}

The two-qubit singlet is the minimal building block of a larger entangled sensor network. Extending the interaction-mediated entanglement to an array of electrons on helium, with engineered spin couplings between neighbors, would prepare a many-body entangled state whose collective response to a localized perturbation is enhanced. A local event affecting one or a few electrons then imprints a correlated signature across the network, detectable through suitable joint observables. This parallels proposals for scalable quantum-dot sensor networks, but the helium platform offers potentially longer coherence times and greater uniformity. The framework thus bridges many-body entanglement theory with practical quantum-enhanced particle detection, in line with the goals of the CERN DRD5 initiative \cite{drd52024}.

The two coupling channels identified above are complementary metrological resources, and a single platform can exploit both. The singlet is the optimal probe for field \emph{gradients} (generator $G_-$), being immune to common-mode drift, whereas Bell- and GHZ-type superpositions are optimal for \emph{common-mode} fields (generator $G_+$); each reaches Heisenberg scaling, $F_Q\propto N^2$, for its respective generator. The unifying quantity is the quantum Fisher information, and the unifying physical resource is the variance of the field-coupling generator in the entangled probe. Engineering networks that host both kinds of probe would let one platform target differential signals (e.g.\ a particle traversing one trap) and global signals (e.g.\ a weak oscillating field from an axion-like background) alike, providing a complete route from the microscopic Hamiltonian to a metrological sensitivity bound.

\section{Discussions, proposed implementation and considerations}

Electrons on superfluid helium are confined by microfabricated electrostatic traps above the liquid surface.  In practice, single electrons are held in lithographically defined potential wells and coupled via their mutual Coulomb interaction.  For example, two electrons in neighboring traps interact through the Coulomb potential (which may be altered due to screening effects)
$V_C = \frac{e^2}{4\pi\epsilon_0|\mathbf{r}_1-\mathbf{r}_2|}$,
giving rise to entangled motional states.  Alternatively, a common superconducting microwave resonator mode can mediate an effective interaction between spatially separated electrons, enabling cavity-based entangling gates.  In both cases, electrostatic control allows tuning of the trap potentials so that specified motional (orbital) levels are brought into resonance; the Coulomb coupling then hybridizes the two-electron states into entangled eigenstates.  The net result is the preparation of two-electron Bell-like states (for instance $|\Psi^\pm\rangle\propto|01\rangle\pm|10\rangle$ in the motional basis), without relying on exchange as in semiconductor dots.

\subsection{Sensing mode}

Once entangled, the electron pair operates in a {\it sensing mode}: external perturbations (such as magnetic or electric fields, or passing charged particles) impart correlated shifts to the joint quantum state.  For example, in a magnetic field $B(t)$ the Zeeman Hamiltonian $H_Z = g_e\mu_B B(t)\,S_z$ causes a phase $\phi = g_e\mu_B\int B\,dt/\hbar$ to accumulate on spin-dependent components of the state.  Similarly, an external electric field or force can shift the orbital frequencies of each trapped electron.  Because the electrons are entangled, these phase shifts translate into a measurable change of the collective state, for example as a relative phase between $|01\rangle$ and $|10\rangle$.  By appropriate choice of entangled input states, common-mode noise can be rejected and sensitivity to the desired signal is enhanced (up to the Heisenberg limit of $\sim1/N$ for $N$ entangled electrons).  In practice one performs a Ramsey-type sequence or parity measurement on the entangled state, converting the acquired phase into a population difference that encodes the signal.

\subsection{Readout}
The readout of electrons-on-helium qubits leverages microwave cavity techniques and image-charge sensing.  In recent experiments, the electron trap is integrated with a superconducting coplanar resonator: the presence of an electron shifts the resonator’s frequency and induces vacuum Rabi splittings when on resonance~\cite{koolstra2026strong}.  By measuring the microwave transmission or frequency shift of the resonator, one can non-destructively infer the electron’s motional state or charge occupancy (even distinguishing 0,1,2,… electrons in the dot~\cite{castoria2025}).  Moreover, transitions between the electron’s quantized vertical (Rydberg) states can be  detected via sensitive image-charge amplifiers: when an electron is excited between Rydberg levels, the change in its image charge induces a current pulse that can be sensed by a cryogenic amplifier.  These methods enable single-shot readout of the electron’s state.  In principle, the electron spin can also be read out by first mapping the spin onto the orbital degree of freedom (e.g.\ via a magnetic field gradient), and then using the above charge/motion readout techniques.

\subsection{Integration into high-energy experiments}

The electrons-on-helium sensor is inherently compatible with cryogenic, low-noise environments typical of particle physics experiments.  The entire device (silicon chip, electrodes and resonator) can be placed inside a dilution refrigerator or a cryogenic superconducting magnet with minimal modification.  Superfluid helium itself is a cryogenic medium (below $\sim0.5$~K) and requires an ultra-low-vibration vacuum space, conditions that many detectors (e.g.\ superconducting qubit arrays or dark matter cavities) already maintain.  The planar microfabricated architecture allows tiling of many traps on a chip, while readout lines and control electrodes can be routed out similarly to other cryogenic sensors.  Importantly, the electron qubits reside above a defect-free liquid interface, free of stray charges or nuclear spins, which means they are intrinsically radiation-hard compared to solid-state qubits.  The predicted coherence times are exceptionally long (orbital coherence well above milliseconds and spin coherence estimated $\gtrsim10$ seconds), ensuring that the sensor maintains quantum coherence over long interrogation periods.  These features make the platform well-suited for integration into high-energy physics experiments where ultralow noise and high sensitivity are required.

\subsection{Challenges and considerations}

\begin{itemize}
\item {\bf Cryogenics and Infrastructure:} The system requires operation at millikelvin temperatures and a supply of superfluid helium.  This adds complexity (cryogen handling, vacuum maintenance) but is similar to requirements for many existing quantum or dark-matter experiments. Additionally, thermal stability and vibration isolation are critical. Recent work has developed methods for vastly simplifying the infrastructure needed for helium filling as well as vibrations produced by thermoacoustic motion of the superfluid~\cite{castoria2024hermetic}. Additionally, dispersive charge sensing techniques based on cQED have been demonstrated to work at temperatures as high as 1~K~\cite{castoria2025}.
\item {\bf Single-Electron Control:} Trapping and manipulating individual electrons on helium is technically challenging, but readily performed.  Precise voltage control is needed to load and isolate single electrons and has been demonstrated using both resonator-based and CMOS-based devices~\cite{castoria2025,castoria2025selective}. While strong microwave photon-electron charge qubit coupling has been achieved~\cite{koolstra2026strong}, work remains to demonstrate single electron spin readout and coherent control, though the initial experiments coupling to the charge qubit state are promising.
\item {\bf Readout Fidelity:} Although strong coupling to resonators has been demonstrated, achieving high-fidelity, quantum-limited readout of spin or motion is still an active area, especially for the spin state of the electron.  Detection of the lateral motional quantum states has been shown, but translating these into fast, low-noise measurements competitive with other qubit platforms requires further development of the readout chain, including the integration of quantum-limited amplifiers such as traveling-wave parametric amplifiers (TWPAs) or Josephson parametric amplifiers (JPAs). Spin state readout presents a particular bottleneck, as the small magnetic moment of a free electron makes direct dispersive coupling to a resonator intrinsically weak, and any spin-to-charge or spin-to-photon conversion step introduces additional infidelity. In this regard, spin-to-charge hybridization in a double-well geometry offers a promising pathway, where the spin degree of freedom is mapped onto a charge state that can then be read out dispersively with significantly stronger coupling~\cite{burkard2023semiconductor}. It remains an open question whether such schemes can be implemented in a quantum non-demolition (QND) fashion on the eHe platform, which is a prerequisite for repeated measurement and, ultimately, quantum error correction. Taken together, these considerations point to readout as one of the most critical near-term development areas for the eHe qubit platform, alongside advances in coherence and gate fidelity.
\item {\bf Fabrication and Scalability:} Microfabricated helium devices involve microchannels to contain the liquid.  Ensuring uniform operation across hundreds, thousands, and ultimately millions of devices, while preserving coherence, will require engineering and circuit layout to minimize cross-talk of the superconducting resonator-based readout devices. The choice of superconducting materials and surface treatments will be critical, as TLS defects, surface losses, and quasiparticle poisoning degrade resonator quality factors and must be mitigated in the helium-wetted cryogenic environment. At the device level, lithographic precision in defining gate electrode geometries directly sets the uniformity of the electrostatic trapping potentials across an array, and disorder in these potentials, whether from fabrication imperfections or charge noise at dielectric interfaces, will limit coherence and gate fidelity at scale.  
\item {\bf Signal Coupling:} The sensor must interact sufficiently with the high-energy particle or field of interest without being shielded by its encapsulation. The appropriate coupling scheme depends critically on the physical quantity being sensed, i.e. electromagnetic, magnetic, acoustic, or otherwise. The challenges arises from the fact that the superconducting and cryogenic enclosure necessary to protect the qubit from noise also screens many of the signals of interest, particularly low-frequency magnetic fields expelled by the Meissner effect. Designing controlled normal-metal breaks that admit the target signal while preserving the low-noise environment is therefore an important electromagnetic engineering consideration. Strong external coupling also risks introducing back-action or excess noise from the signal channel into the sensor, motivating the development of on-chip isolation strategies compatible with the eHe environment. Taken together, these considerations suggest that signal coupling is not merely a peripheral engineering detail but a co-design problem that must be addressed alongside qubit coherence and readout from the earliest stages of sensor development.
%Designing coupling schemes (e.g.\ integrating magnets or antennas) that bring the target signal to the electron qubits is an important consideration.
\end{itemize}

Despite these challenges, the electrons-on-helium platform offers unique advantages (no nuclear spins, and long coherence) that make it a promising candidate for entangled quantum sensing in particle physics applications.  Continued advances in cryogenic control and circuit-QED integration are steadily improving its viability as a quantum sensor.

\section{Conclusions and perspectives}

We have outlined a conceptual and theoretical framework for a quantum sensor based on spin-entangled electrons on helium, aligned with the goals of the CERN DRD5 initiative \cite{drd52024} on quantum sensing for particle physics. By using entangled electron spins trapped above a superfluid helium surface as sensitive probes, the proposed sensor aims to use quantum entanglement to optimize the detection of particle physics events. In principle, this approach could outperform classical detection methods in certain regimes.  For comparison, similar ideas have been proposed using spin-entangled semiconductor quantum dots (QDs), but electrons on helium offer unique advantages: the electrons reside in a near-pristine environment (free of crystal defects or nuclear spins) that promises exceptionally long coherence times and high mobility. These features make the electron-on-helium platform particularly well suited to maintaining entanglement and achieving high sensitivity.

We described how a pair of entangled electrons on helium (for example in a spin-singlet state) can act as a precise detector of differential perturbations, and how this principle could be extended to larger networks of sensors for enhanced sensitivity. The extreme isolation of electrons on helium means that tiny external fields or forces (for instance from passing particles or dark matter) could induce measurable changes in their joint quantum state. Building a network of such entangled sensors could enable differential measurements with unprecedented precision.

Integrating these sensors into particle physics experiments could open new ways to observe rare processes and subtle signals. For example, spin-entangled helium electron sensors might be used to search for physics beyond the Standard Model by detecting very small field effects or improving the resolution of energy deposition profiles in detectors. Because electrons on helium can be localized in micro-fabricated traps and precisely manipulated, this approach could potentially complement or enhance traditional detector elements. In practice, one might imagine deploying arrays of helium-based quantum sensors at strategic locations in an experiment to pick up minute signals that would be too small for conventional devices.

This proposal is inherently interdisciplinary, sitting at the intersection of quantum information science and high-energy physics instrumentation. The theoretical motivation is strong: quantum metrology predicts significant gains from entanglement-enhanced sensing, and the platform of electrons on helium has been extensively studied in the quantum computing community. Early theoretical studies and experimental efforts on electrons trapped above liquid helium have demonstrated that single electrons can be controlled and entangled with high fidelity. These results support the feasibility of pursuing quantum-enhanced sensors in this system. While semiconductor quantum dots represent a more mature technology, the helium-electron system stands out as a promising alternative due to its remarkably clean environment.

However, practical realization of an electrons-on-helium sensor will require overcoming significant challenges. Coherence must be preserved in a realistic experimental setting, and one must engineer traps, readout, and control that work reliably under detector conditions. We emphasize that the DRD5/RD quantum program \cite{drd52024} is an ideal incubator for such high-risk, high-reward ideas. By pursuing research on helium surface materials, cryogenic electron trapping techniques, entanglement generation methods, and robust sensor design, the particle physics community can assess whether the promised quantum advantage can be harnessed in future detectors.

In closing, the spin-entangled electron-on-helium sensor stands as a conceptual example of how thinking quantum-mechanically about detectors could revolutionize the way we measure events in particle physics. It invites both theorists and experimentalists to collaborate: theorists can refine the models (for instance by including realistic noise sources and the detailed interaction of the sensor with specific physics processes and the environment), while experimental teams can begin testing components (such as single-electron traps on helium or schemes to generate spin entanglement in situ). If successful, the outcome would be a new class of detector technology that adds a fundamentally quantum toolkit to the exploration of fundamental physics, complementing conventional approaches and offering sensitivity to effects otherwise inaccessible.

\medskip
\noindent \textbf{Data availability :}
All of the data produced in association with this work have been stored and are publicly available  at \url{https://github.com/mariaelenaUiA-hub/Double-Well_RL}. The data presented in the article can be reproduced using the codes from the above repository. 

\medskip
\noindent \textbf{Code availability :}
All codes (full configuration interaction theory, reinforcement learning and data analysis) developed in association with this work have been stored and are publicly available  at \url{https://github.com/mariaelenaUiA-hub/Double-Well_RL}.

\begin{acknowledgments}
The Michigan State University (MSU) portion of this work was supported by the U.S. National Science Foundation via grant number DMR-2410650 and the Cowen Family Endowment at MSU. G.F.L. and V.S. acknowledge funding from the European Union’s Horizon Europe research and innovation program under the Marie
Skłodowska-Curie Grant Agreement No. 101126636. O.L. has received funding from the European Union's Horizon 2020 research and
innovation program under the Marie Skłodowska-Curie grant agreement No. 945371. M.H.-J. is partially supported by U.S. National Science Foundation (NSF) Grants PHY-1404159 and PHY-2310020. A.C., J.M. and H.S. acknowledge funding from the Research Council of Norway under grant agreement 310713 (NorCC center). F. P. M. acknowledges financial support from the Research Council of
Norway (Grant No. 333937) through participation in the QuantERA ERA-NET Cofund in Quantum Technologies.
\end{acknowledgments}

\medskip
\noindent \textbf{Author contributions:}
S.D.B., J.B.F., M.H.-J., G.F.L., O.L., F.P.M., M.E.P., V.S., Z.J.S., and J.D.W. carried out the theoretical and numerical analyses. M.E.P. developed the reinforcement learning codes and analysis tools. N.R.B. and J.P.  developed and contributed to the experimental  analyses of electrons on helium. A.Y.M.C.C., J.M., and H.S. contributed to the discussions of the DRD5 project of CERN. M.H.-J. and F.P.M. supervised the work. All authors contributed to the discussion of the results and the manuscript.

\medskip
\noindent \textbf{Competing interests :}
The authors declare no competing interests.

% Some references carry a `language' field; apsrev4-2 then emits \selectlanguage
% in the .bbl. Neutralize it here so that babel is not required.
\makeatletter
\def\selectlanguage#1{}
\makeatother
\bibliographystyle{apsrev4-2}
%apsrev4-2.bst 2019-01-14 (MD) hand-edited version of apsrev4-1.bst
%Control: key (0)
%Control: author (72) initials jnrlst
%Control: editor formatted (1) identically to author
%Control: production of article title (-1) disabled
%Control: page (0) single
%Control: year (1) truncated
%Control: production of eprint (0) enabled
%

%\bibliography{references}

\end{document}